\documentclass[aps,prb,twocolumn,amsmath,amssymb,showpacs,superscriptaddress]{revtex4-2}
\usepackage{graphicx}						
\usepackage{epstopdf}
\usepackage{amsmath,amssymb,amsfonts}
\usepackage{textcomp}
\usepackage{hyperref}
\usepackage{gensymb}
\usepackage{verbatim}
\usepackage{amsmath}
\hypersetup{breaklinks=true,colorlinks=true,urlcolor=black}
\usepackage{color}
\usepackage{soul}
\usepackage{float}
\usepackage{gensymb}
\usepackage{caption}
\usepackage{subcaption}
\usepackage[english]{babel}
\usepackage[T1]{fontenc}
\usepackage[utf8]{inputenc}
\usepackage{booktabs}
\DeclareGraphicsExtensions{.jpg,.pdf,.png,.eps}

\begin{document}
\title{Unusual coercivity and zero field stabilization of fully saturated magnetization in single crystals of LaCrGe$_3$}
\author{M. Xu}
\affiliation{Ames National Laboratory, Iowa State University, Ames, Iowa 50011, USA}
\affiliation{Department of Physics and Astronomy, Iowa State University, Ames, Iowa 50011, USA}

\author{S. L. Bud'ko}
\affiliation{Ames National Laboratory, Iowa State University, Ames, Iowa 50011, USA}
\affiliation{Department of Physics and Astronomy, Iowa State University, Ames, Iowa 50011, USA}

\author{R.~Prozorov}
\affiliation{Ames National Laboratory, Iowa State University, Ames, Iowa 50011, USA}
\affiliation{Department of Physics and Astronomy, Iowa State University, Ames, Iowa 50011, USA}

\author{P. C. Canfield}
\affiliation{Ames National Laboratory, Iowa State University, Ames, Iowa 50011, USA}
\affiliation{Department of Physics and Astronomy, Iowa State University, Ames, Iowa 50011, USA}
\email[]{canfield@ameslab.gov}

\date{\today}

\begin{abstract}
LaCrGe$_3$ is an itinerant, metallic ferromagnet with a Curie temperature ($T_C$) of $\sim$ 86 K.  Whereas LaCrGe$_3$ has been studied extensively as a function of pressure as an example of avoided ferromagnetic quantum criticality, questions about its ambient pressure ordered state remain; specifically, whether there is a change in the nature of the ferromagnetically ordered state below $T_C$ $\sim$ 86 K.  We present anisotropic $M$($H$) isotherms, coupled with anisotropic AC susceptibility data, and demonstrate that LaCrGe$_3$ has a remarkable, low temperature coercivity associated with exceptionally sharp, complete magnetization reversals to and from fully polarized states. This coercivity is temperature dependent, it drops to zero in the 40 - 55 K region and reappears in the 70 - 85 K regions. At low temperatures LaCrGe$_3$ has magnetization loops and behavior that has previously associated with micromagnetic/nanocrystalline materials, not bulk, macroscopic samples.
\end{abstract}

\maketitle

\section{introduction}

Recently, LaCrGe$_3$ has become an intensively studied metallic, itinerant ferromagnet, in particular, due to its complex but comprehensible behavior under pressure. \cite{Taufour2016,Kaluarachchi2017a,Taufour2018,Gati2021,Rana2021,Nguyen2018} In addition to intriguing high pressure properties, the ambient pressure characteristics of LaCrGe$_3$ potentially make it a fertile ground to study different aspects of magnetic behavior.

LaCrGe$_3$ crystallizes in a hexagonal structure (space group $P6_3/mmc$). \cite{Bie2007,Cadogan2012} The samples of LaCrGe$_3$ can be grown in a single crystal form \cite{Lin2013} with a rod-like morphology ($c$ - axis along the rod's axis) and up to several hundreds mg mass. At ambient pressure LaCrGe$_3$ is a ferromagnet with some degree of itinerancy, with a Curie temperature of $T_C$ $\sim 86$~K, and magnetic moments ordered along the $c$-axis with the value of the saturated moment, $\mu_S \approx 1.2$~$\mu_B$/Cr \cite{Bie2007,Cadogan2012,Lin2013,Bosch-Santos2021} and an anisotropy field of 40 - 50 kOe. \cite{Lin2013,Taufour2018} Although the ground state of LaCrGe$_3$ appears to be that of a simple ferromagnet, there are several experimental observations that require further consideration and understanding. The low field, temperature dependent magnetization, measured using field-cooled protocol, has an unusual shape, exhibiting a small peak around 68 K. \cite{Lin2013}  Electrical transport measurements suggested two ferromagnetic phase transitions with $T_C \sim 86$~K and $T_x \sim 71$~K,\cite{Taufour2018} although no corresponding features were observed in thermodynamic measurements. \cite{Taufour2016,Lin2013} Electron spin resonance \cite{Sichelschmidt2021} and AC susceptibility \cite{Bosch-Santos2021} measurements on polycrystalline samples reveal, in addition to $T_C$, some anomalies in the 40 - 50~K temperature range. All these features are not understood and strongly suggest the need for more detailed studies of the temperature and field dependent magnetization of LaCrGe$_3$.

In order to better identify possible changes in the ferromagnetically ordered state, we performed systematic, anisotropic magnetization measurements. As a result we have found that for field applied along the $c$-axis, LaCrGe$_3$ manifests exceptionally sharp, square, hysteretic, magnetization loops that imply that there is a discontinuous reversal of the fully saturated magnetization of a bulk sample at a well defined field. Of many possible phenomena associated with ferromagnetism, the coherent magnetization reversal of an uniformly magnetized (single domain) sample, is interesting for an apparent simplicity of the model that describes it \cite{E.C.Stoner1948} as well as for potential applications. In all so far known cases, the monodomain regime is found in small particles, typically below 100 nm in size, consistent with the theoretical consideration of the energy balance between the energy of stray fields (proportional to particle's volume) around such particle and the energy of the domain wall formation ( proportional to the cross-sectional area) \cite{Brown1968,DiFratta2012}. In macroscopic, bulk samples magnetic domains always form. In this paper we will show that bulk, single crystalline LaCrGe$_3$ appears to be a macroscopic manifestation of such a zero field, fully polarized, state that undergoes a discontinuous and full reversal of magnetization at a well defined, finite, coercive field.

\section{Crystal Growth and Experimental Method}
Single crystals of LaCrGe$_3$ were grown in two steps from melts of La$_{18}$Cr$_{12}$Ge$_{70}$ \cite{Canfield2020,Slade2022} using fritted Canfield Crucible Sets (CCS).\cite{Canfield2016c,INC} We first heated the La$_{18}$Cr$_{12}$Ge$_{70}$ to 1150 $\degree$C and cooled to 950 $\degree$C over 50 – 100 hours.  At 950 $\degree$C, a mixture of LaGe$_{2-x}$ plates and LaCrGe$_3$ rods were separated from the remaining liquid phase. This decanted liquid was subsequently resealed, heated to 1000 $\degree$C (so as to fully remelt it) and then slowly cooled from 950 $\degree$C to 825 $\degree$C over roughly 100 hours.  At 825 $\degree$C the growth was decanted and the resulting single phase LaCrGe$_3$ crystalline solid phase was separated from excess liquid.  These crystals were used in our study. All of the LaCrGe$_3$ data presented in this paper, with the exception of figures \ref{22223XRD} and \ref{PDd} in the appendix, were taken on a specific single crystal of LaCrGe$_3$ shown in the inset of figure \ref{22223MTfig1}. Similar results are found for other single crystalline samples as discussed below and shown in figure \ref{PDd}, in the appendix.

LaCrGe$_3$ forms as metallic, rod-like, brittle single crystals. Powder x-ray diffraction measurements were made by grinding single crystals in an agate mortar and pestle and placing the powder onto a single crystalline silicon, zero background sample holder with a small amount of vacuum grease. Powder X-ray diffraction measurements were carried out by using a Rigaku MiniFlex II powder diffactometer in Bragg-Brentano geometry with Cu K$\alpha$ radiation ($\lambda$ = 1.5406 \AA{}). The result of powder X-ray diffraction measurement agrees well with literature \cite{Bie2007} and is shown figure \ref{22223XRD}, in the appendix.

Temperature- and magnetic-field-dependent DC and VSM magnetization data were collected using Quantum Design (QD), Magnetic Property Measurement Systems (MPMS and MPMS3). Temperature- and magnetic-field-dependent DC magnetization measurements were taken for \textit{H} parallel and perpendicular to the crystallographic \textit{c}-axis by placing the rod-like sample between two collapsed plastic straws with the third, uncollapsed, straw providing support as a sheath on the outside or by using of a quartz sample holder. Samples were fixed on the straw or quartz sample holder by GE-7031-varnish. In VSM magnetization measurements a peak amplitude of 4 mm and an averaging of time 2 sec were used. The sample for VSM was glued to a quartz sample holder by GE-7031-varnish. AC susceptibility measurements were performed using the AC-option of the MPMS3 magnetometer for two orientations of the LaGrGe$_3$ crystal and for two frequencies, 7.57 Hz and 75.7 Hz, in 5 Oe ac field and zero dc bias field (after demagnetization procedure was applied to the superconducting magnet). These measurements were performed on cooling. For some of the magnetization measurements made we needed to set a finite applied magnetic field to zero at base temperature without demagnetization; the remnant field of the superconductor coil for these measurements was measured by lead (285.2 mg, 6-9's grade Cominco American) and palladium (QD standard, 258.2 mg, Serial number: 18060702-058) in both the MPMS and MPMS3 units used. The remnant field value in the MPMS3 70 kOe magnet after setting the field to zero at base temperature from + 20 kOe was $\sim$ -19 Oe.  The remnant field for the MPMS 55 kOe magnet after setting the field to zero at base temperature from + 20 kOe was $\sim$ -3 Oe. These remnant field values will be important when we discuss figures \ref{22223TrapF} and \ref{22223TrapFdF} below.

\section{Magnetization Measurements}

\begin{figure}	
	\begin{minipage}{0.45\textwidth}
		\centering
		\includegraphics[width=1.5\columnwidth]{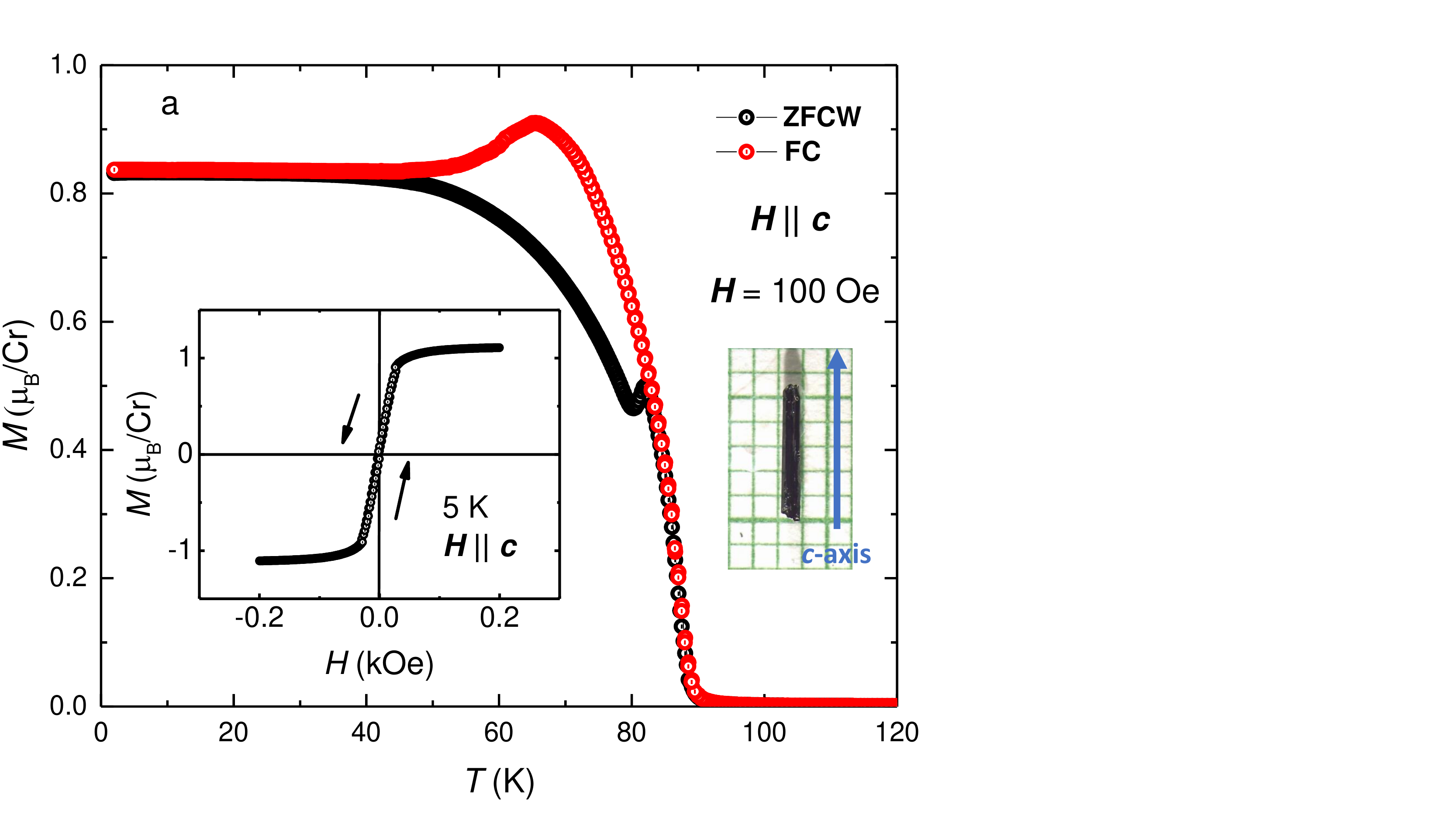}	
	\end{minipage}\hfill
\begin{minipage}{0.45\textwidth}
	\centering
	\includegraphics[width=1.5\columnwidth]{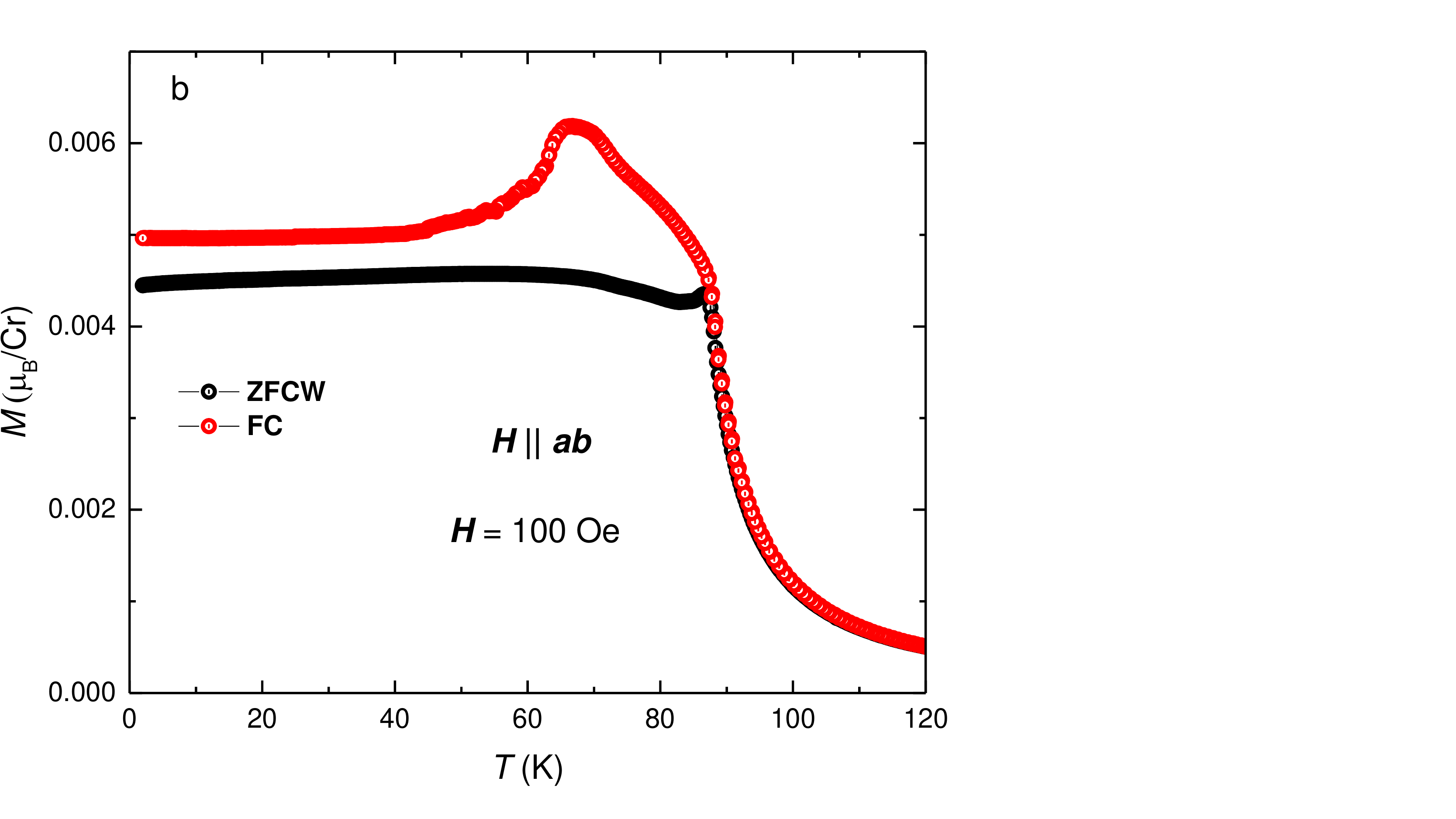}	
\end{minipage}\hfill
	\caption{Zero-field-cooled-warming (ZFCW) and field-cooled (FC) low temperature magnetization as a function of temperature for the LaCrGe$_3$ single crystal with a field of 100~Oe applied parallel (a) or perpendicular (b) to the crystallographic \textit{c}-axis. Left inset shows 4 quadrants of magnetization at 5 K as a function of magnetic field applied parallel to the crystallographic \textit{c}-axis with a very small hysteresis and the coercive field is $\sim$ 3 Oe(not resolvable on this scale). Right inset shows the picture of measured LaCrGe$_3$ single crystal with blue arrow denoting crystallographic \textit{c}-axis.}\label{22223MTfig1}
\end{figure}

Figure \ref{22223MTfig1}a shows the low temperature (1.8~K - 120~K), zero-field-cooled-warming (ZFCW) and field-cooled (FC) magnetization of the rod-like LaCrGe$_3$ single crystal(see inset) with 100 Oe magnetic field parallel to the crystallographic \textit{c}-axis (a) and perpendicular to the $c$-axis (i.e. $H$ || $ab$).  (Figure \ref{22223MTA} in the appendix shows similar data for applied fields of 25 Oe, 50 Oe, and 100 Oe for comparison.) For $H$ || $c$, the ferromagnetic transition, $T_C$ $\sim$ 86 K, and the peak around 70 K are similar to what was found at 50 Oe in reference \cite{Lin2013}. Whereas the kink like feature in the ZFCW data is not unusual, often associated with domain wall motion upon warming, the feature near 70 K in the FC data that leads to a decrease of $M$($T$) on further cooling is somewhat unusual. The inset of figure \ref{22223MTfig1} shows a 5 K, 4-quadrant $M$($H$) loop for fields applied parallel to the crystallographic $c$-axis up to 0.2 kOe. (Field increases from 0 Oe to 0.2 kOe, then decreases to -0.2 kOe and then returns to 0 kOe.) There is a very small hysteresis and the coercive field is $\sim$ 3 Oe. Such a small hysteresis is not unusual for single crystalline ferromagnetic samples which often have very small bulk pinning. For $H$ || $ab$ there is a much smaller response, given that $c$-axis is the axis of strong unidirectional anisotropy and $H$ || $c$ is the easy direction of the ferromagnetically ordered state.

\begin{figure}
	\includegraphics[width=1.5\columnwidth]{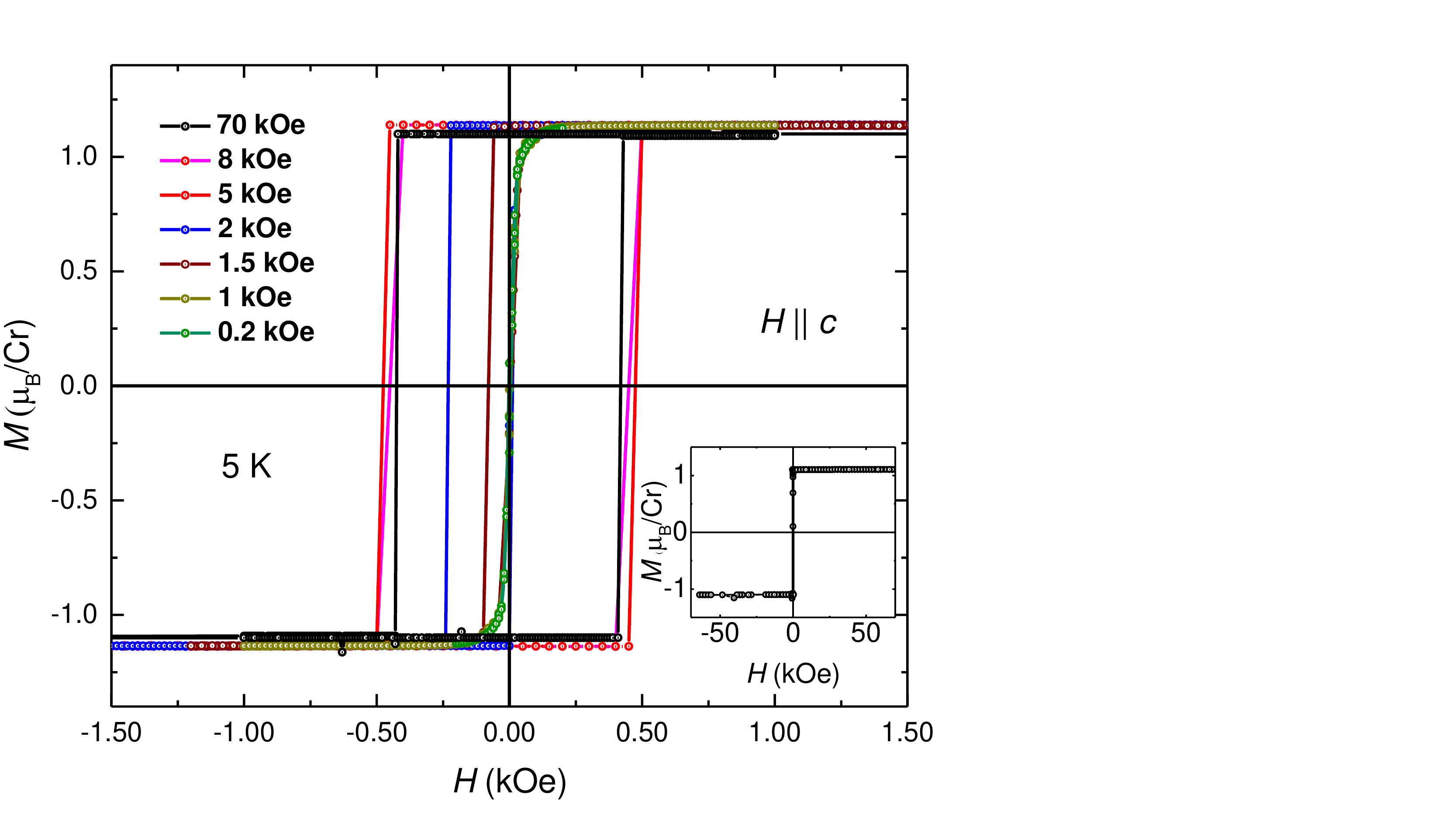}	
	\caption{Magnetization of a LaCrGe$_3$ single crystal at 5 K as a function of magnetic field applied parallel to the crystallographic \textit{c}-axis. We demagnetized the system at 120 K, zero field cooled to 5 K, applied a given maximum field denoted by different color along the $c$-axis. Inset shows $H_{max}$ = 70 kOe $M$($H$). }\label{2223MHdH}
\end{figure}

To our surprise, we found that much larger and sharper(essentially vertical, discontinuous jumps) hysteresis exists for LaCrGe$_3$ for larger, H $\geq$ 5 kOe, applied fields. In figure \ref{2223MHdH} we show, 5 K, $M$($H$) loops for systematically increasing maximum applied field. For the data in each plot we demagnetized the system at 120 K (much higher than T$_C$ = 86 K), zero field cooled to 5 K, applied a given maximum field along the $c$-axis, and then collected a 4-quadrant ($H$$_{max}$ to -$H$$_{max}$ to $H$$_{max}$) $M$($H$) curve. For maximum fields of 0.2 kOe and 1.0 kOe there are essentially reversible $M$($H$) plots. For maximum applied fields of 1.5, and 2.0 kOe there are discontinuously sharp and increasingly hysteretic $M$($H$) plots and for maximum applied fields of 5.0 kOe and greater the coercive field saturates near $\sim$0.5 kOe. For all of these $M$($H$) curves, the magnetization saturates near 1.1~$\mu_B$/Cr. This type of $M$($H$) curve, is very unusual for a bulk, macroscopic, single crystal. The fact that the system has full magnetization persevered for $\sim$ $\pm$0.5 kOe and then discontinuously switches to the same full magnetization with the opposite sign is more commonly seen for nano-crystals rather than bulk samples with dimensions measured in mm.

Such rectangular-looking magnetization loops are found in the phenomenological Stoner-Wohlfarth model for the magnetic field parallel to the anisotropy easy axis \cite{Brown1968,DiFratta2012}. The similar loops are also obtained in direct micromagnetic simulations \cite{Kuzma2022} and analytical theories \cite{Valdes2021}. Experimentally such $M$($H$) loops are observed in elongated (shape anisotropy) and/or magneto-crystalline-anisotropic nanoparticles \cite{Trung2016}. However, only at low temperatures, below so-called blocking temperature \cite{Prozorov1999a} because above it a particle's magnetic moment can be easily flipped by thermal fluctuations, $k_B$$T$ \cite{Trung2016}. (Indeed, in our case of large crystals thermal activation is not relevant in a large temperature interval.) Experimentally coherent magnetization reversal was directly observed in 25 nm nanoparticles using magnetic force microscopy(MFM) imaging, but showed that larger particles favor different modes (e.g. so-called vortex core switching) of the reversal \cite{Pinilla-Cienfuegos2016}. In addition to a classical rotation of the Stoner-Wohlfarth type, nanoparticle magnetization reversal can also proceed via quantum tunneling of virtual domain walls; both were studied using $\mu$-SQUID measurements \cite{Wernsdorfer2000}. Of course, in macroscopic, bulk single crystals such as the LaCrGe$_3$ samples studied here, quantum switching in not feasible. In another work also using $\mu$-SQUID, not only was agreement with the Stoner-Wohlfarth model found, but it was also shown that surface pinning plays crucial role in magnetization reversal \cite{Jamet2001}. In the absence of significant bulk pinning such surface pinning contributions may also be relevant for bulk crystals.

\begin{figure}
	\centering
	\begin{minipage}{0.45\textwidth}
		\centering
		\includegraphics[width=1.5\columnwidth]{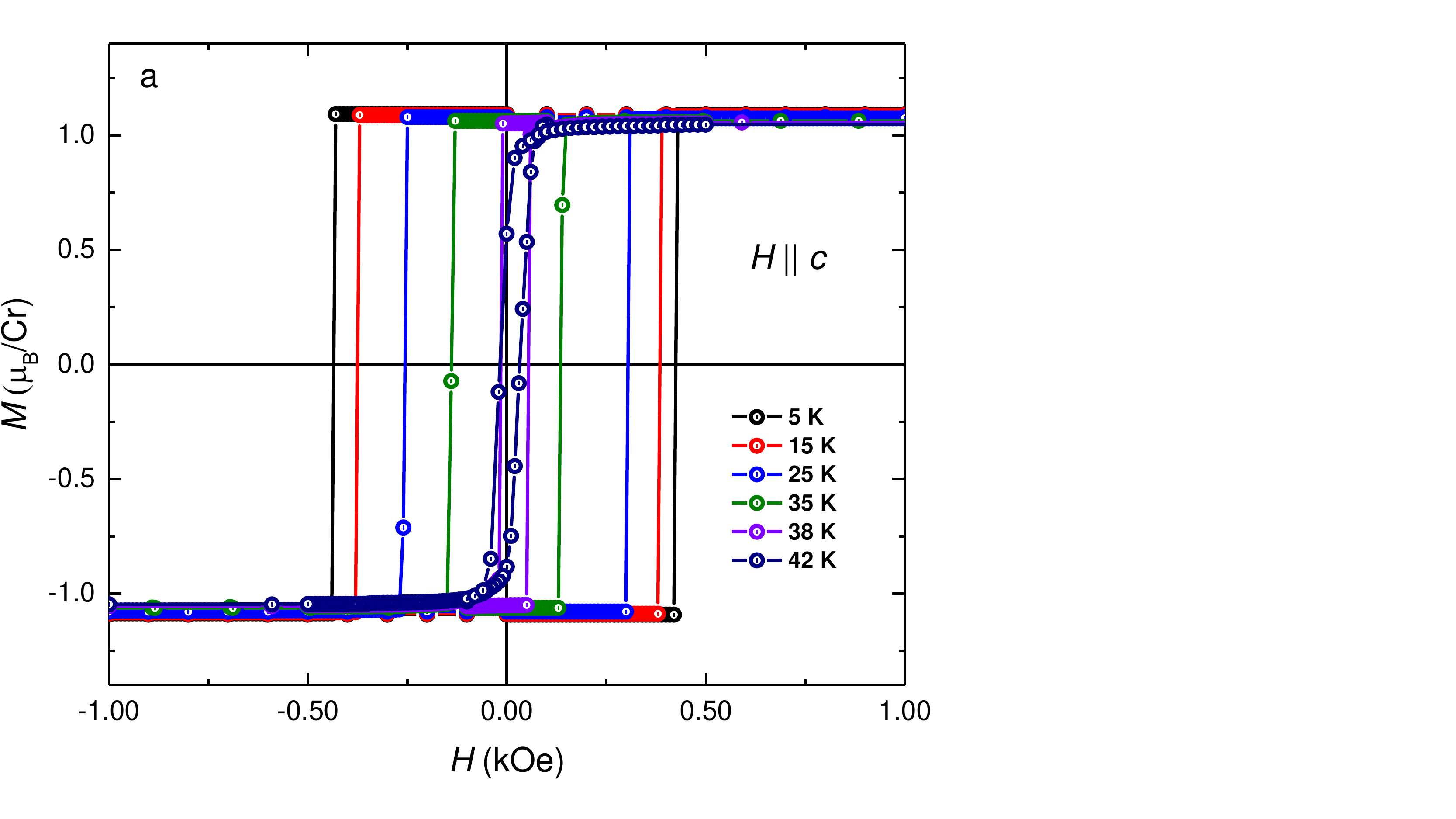}		
	\end{minipage}\hfill
	\centering
	\begin{minipage}{0.45\textwidth}
		\centering
		\includegraphics[width=1.5\columnwidth]{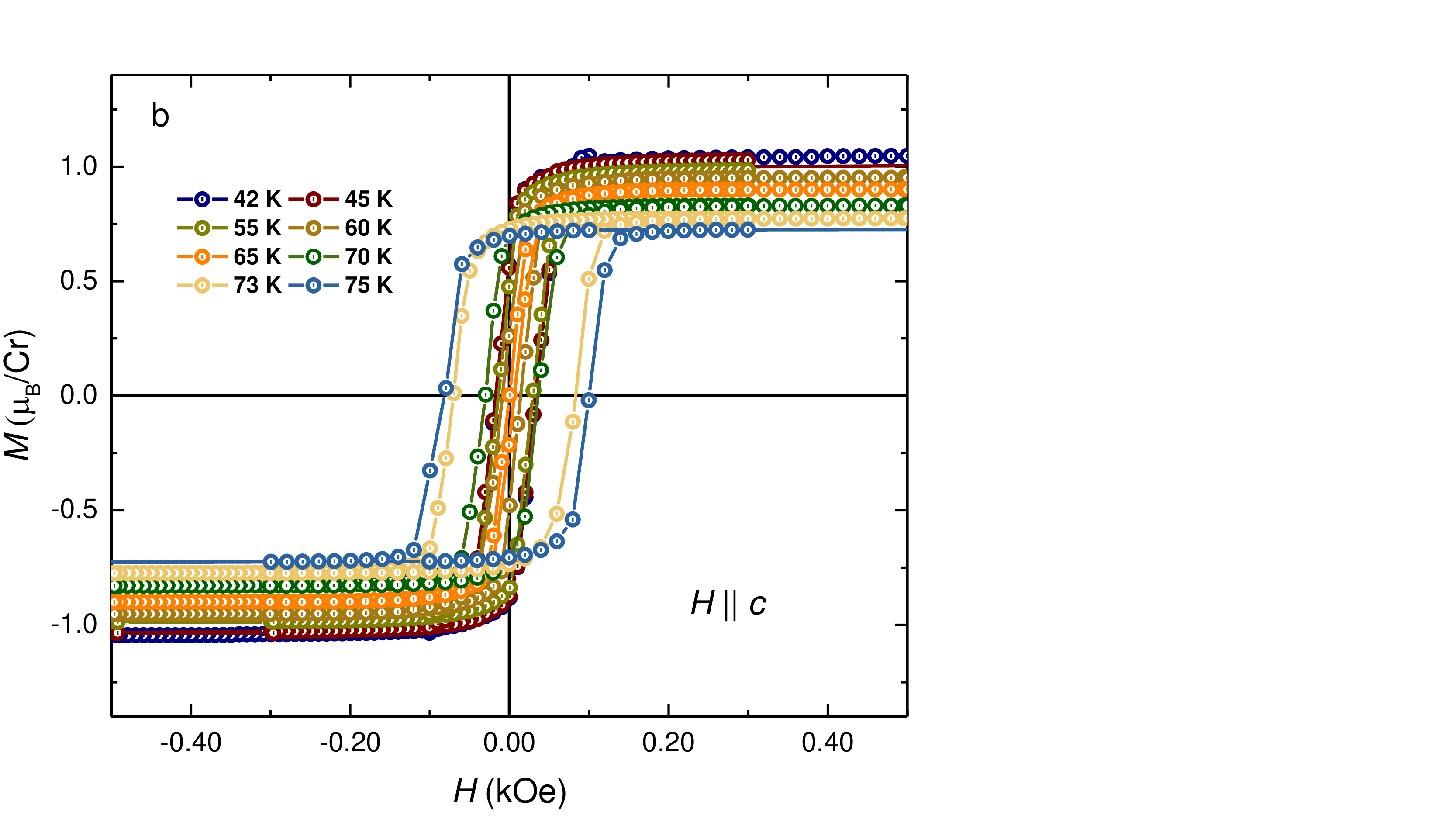}		
	\end{minipage}\hfill
	\centering
	\begin{minipage}{0.45\textwidth}
		\centering
		\includegraphics[width=1.5\columnwidth]{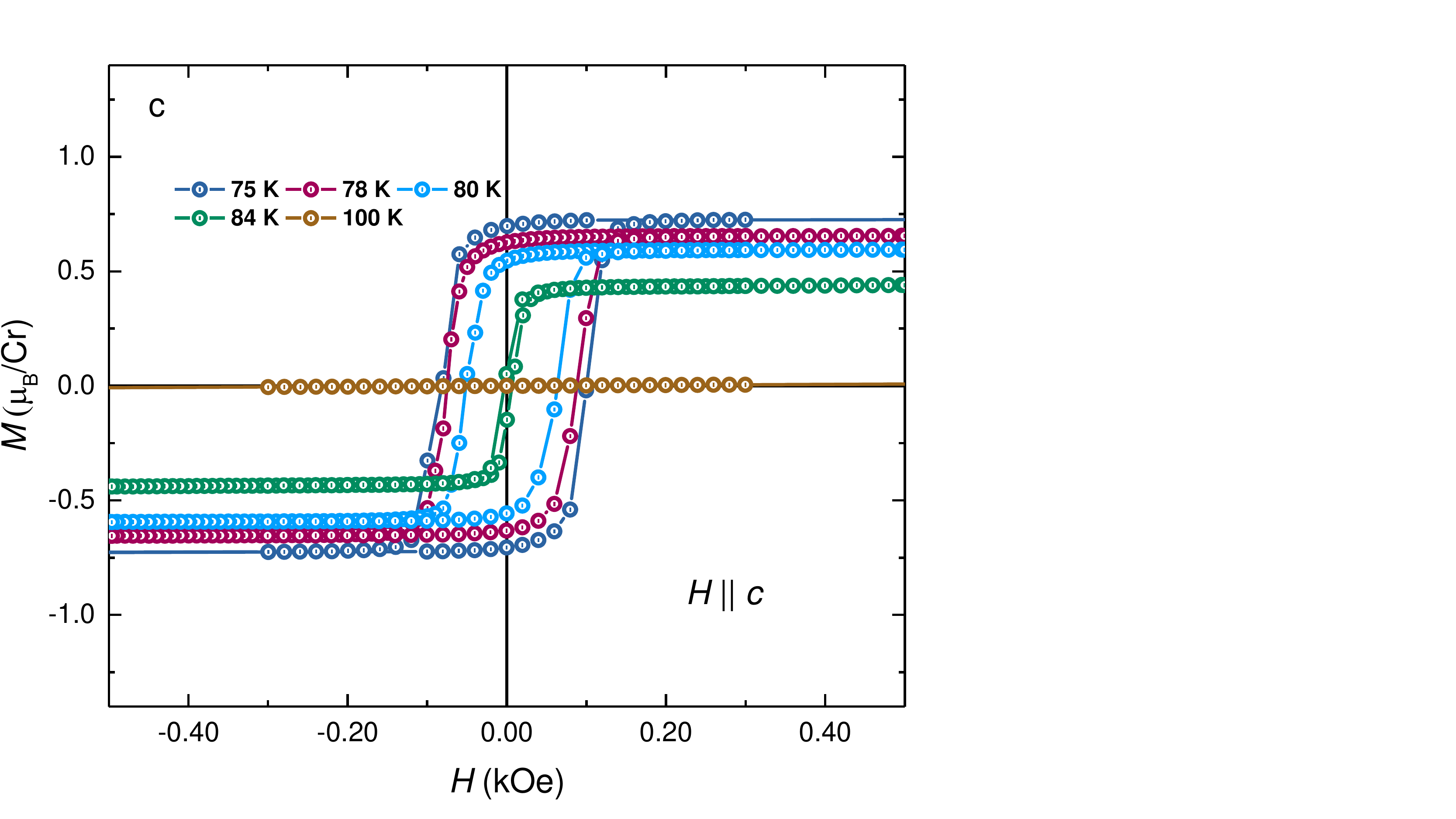}		
	\end{minipage}\hfill
	\caption{Magnetization of a single crystal of LaCrGe$_3$ at different temperature as a function of magnetic field applied parallel to the crystallographic \textit{c}-axis. Each isothermal loop is a 4-quadrant loop with field being swept from +5 kOe to -5 kOe and back to +5 kOe. Between loops the system is taken to 120 K, and then cooled in zero field to the next temperature. a) shows $M$($H$) at $T \leq 42$ K. b) shows $M$($H$) at $42$~K $\leq T \leq 75$~K. c) shows $M$($H$) at $75$~K $\leq T \leq 100$~K. }\label{22223MHdTC}
\end{figure}

In order to study the temperature evolution of LaCrGe$_3$'s coercivity, we measured $M$($H$) isotherms from 5 K to above T$_C$. Figure \ref{22223MHdTC} shows magnetization of a single crystal of LaCrGe$_3$ as a function of magnetic field applied parallel to the crystallographic \textit{c}-axis at different temperatures. The maximum applied field was 5 kOe. In figure \ref{22223MHdTC}a, $M$($H$) behaves similar to the 5 K data in figure \ref{2223MHdH} and coercive fields decrease  monotonically as temperature increase. Ferromagnetism with "softer" behavior is shown at $T = 42$~K. Figure \ref{22223MHdTC}b shows $M(H)$ in the temperature range of $42$~K $\leq T \leq 75$~K. Coercivity is low and almost constant up to $\sim$ 55 K and then starts to increase as temperature increases further. Figure \ref{22223MHdTC}c shows the $M(H)$ loops for $75$~K $\leq T \leq 100$~K. In this temperature range, the coercive field decreases and for 100 K > $T_C$ the response is only weakly paramagnetic. Although there is clear hysteresis visible in the $M$($H$) data in the higher temperature region, it is not as sharp ($M$($H$) loops not as square) as it is for $T$ < 40 K.

\begin{figure}
	\includegraphics[width=1.5\columnwidth]{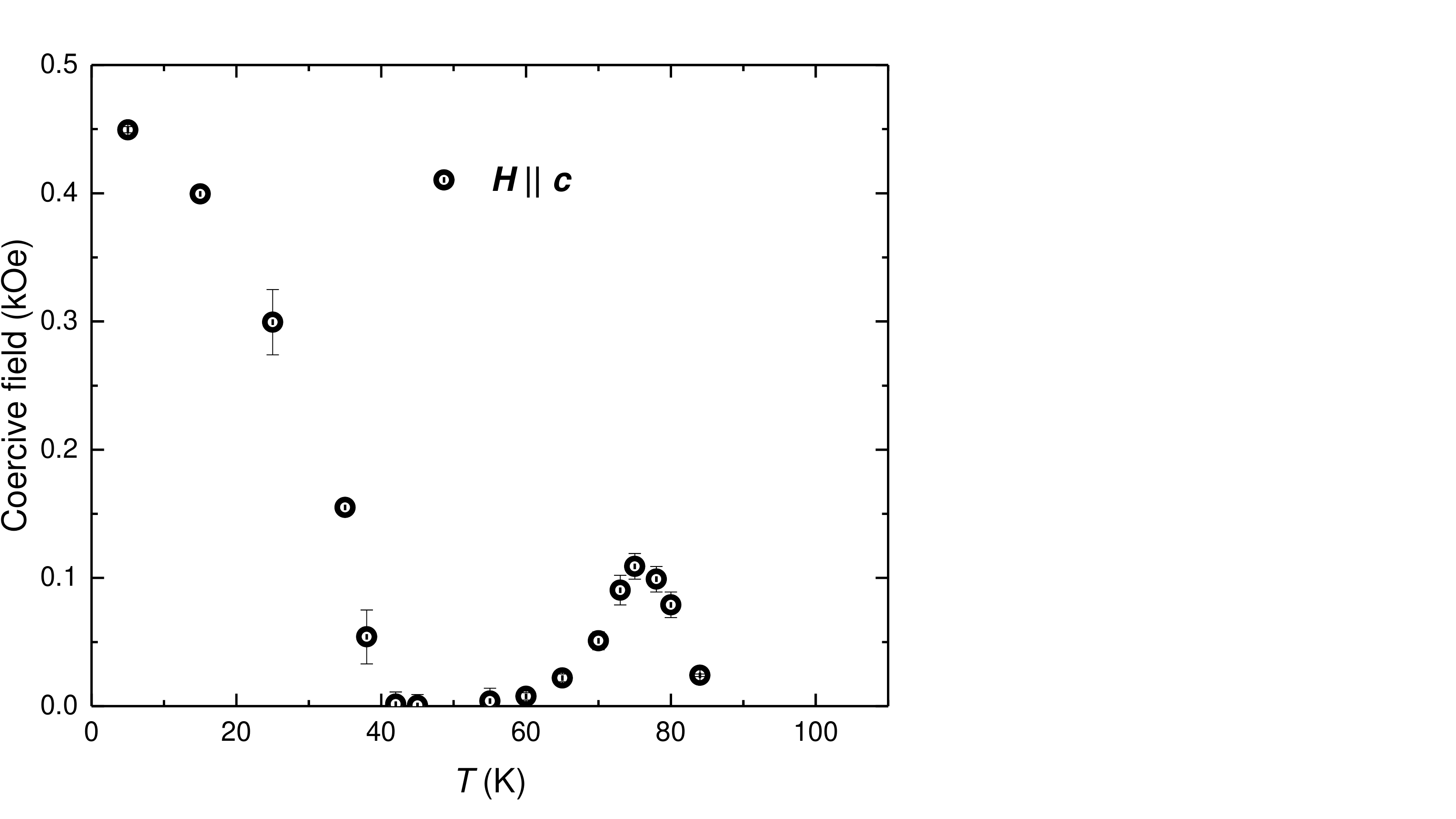}	
	\caption{Coercive field of a LaCrGe$_3$ single crystal as a function of temperature with magnetic field applied parallel to the crystallographic \textit{c}-axis.}\label{PD}
\end{figure}

The coercivity fields inferred from figures \ref{22223MHdTC} are plotted as a function of temperature in figure \ref{PD}, the 5 K coercivity drops to zero just above 40 K, stays near zero up to 60 K and then rises through a local maxima at $\sim$ 75 K and drops to zero at $T_C$ $\sim$ 86 K. We find that different single crystalline samples of LaCrGe$_3$ have qualitatively similar, coercive field as a function of temperature plots, as shown in Appendix figure \ref{PDd}. The primary difference between samples is the size of the 5 K coercive field and the precise temperature it drops to zero near 40-50 K; the higher temperature coercivity seems less variable, sample to sample.  

\begin{figure}
	\includegraphics[width=1.5\columnwidth]{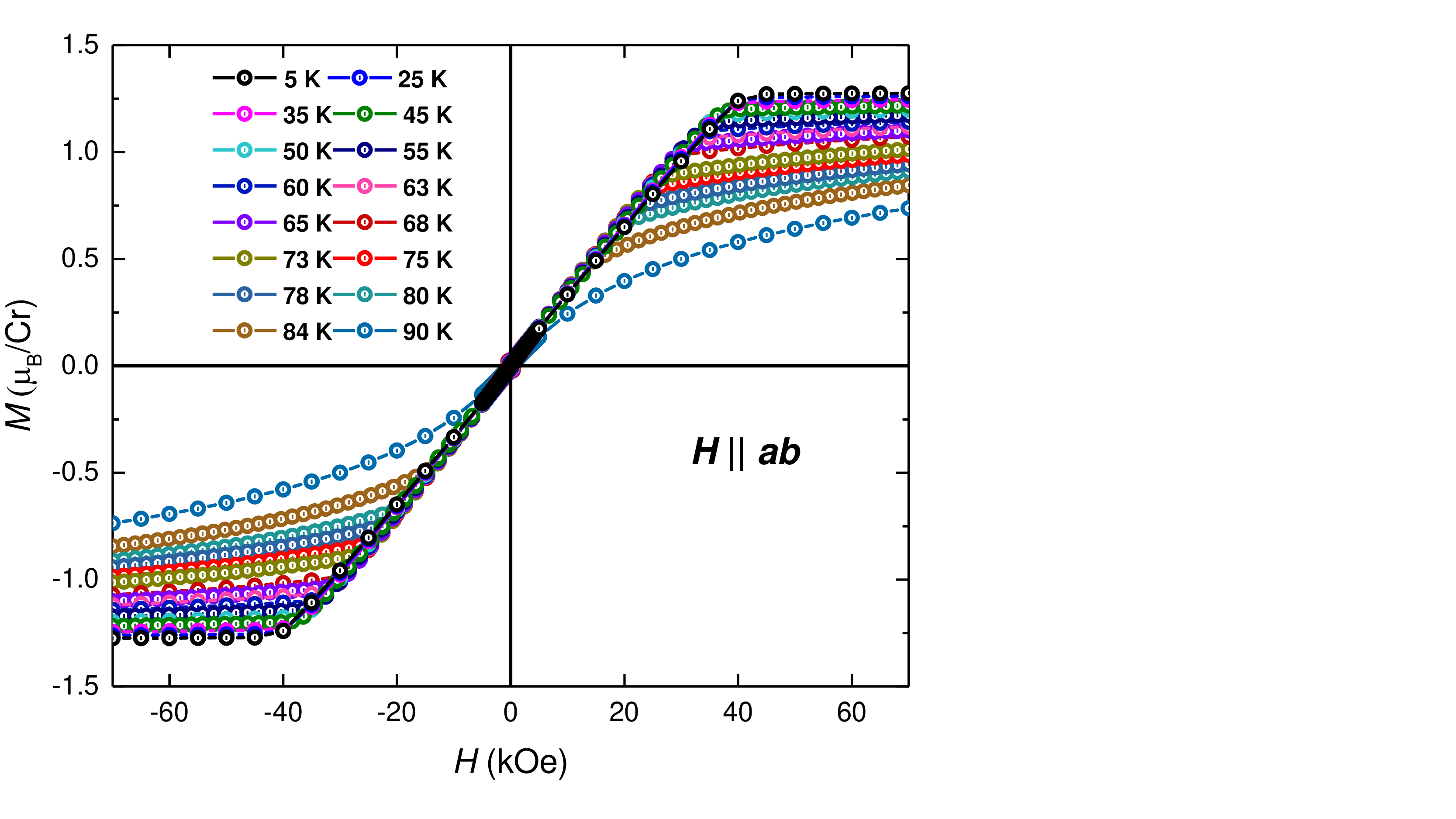}	
	\caption{Magnetization of a LaCrGe$_3$ single crystal at different temperature as a function of magnetic field applied parallel to the crystallographic \textit{ab}-plane.}\label{22223MHdTABF}
\end{figure}

Figure \ref{22223MHdTABF} shows magnetization at different temperatures as a function of magnetic field applied parallel to the crystallographic \textit{ab}-plane. At 5 K, the magnetization saturates for $H$ > 40 kOe; for $H$ < 40 kOe the $M$($H$) plot appears to be linear, at least on the scale shown in figure \ref{22223MHdTABF}. As temperature increases the deviation from the low field, linear behavior decreases.

\begin{figure}
	\centering
	\begin{minipage}{0.45\textwidth}
		\centering
		\includegraphics[width=1.5\columnwidth]{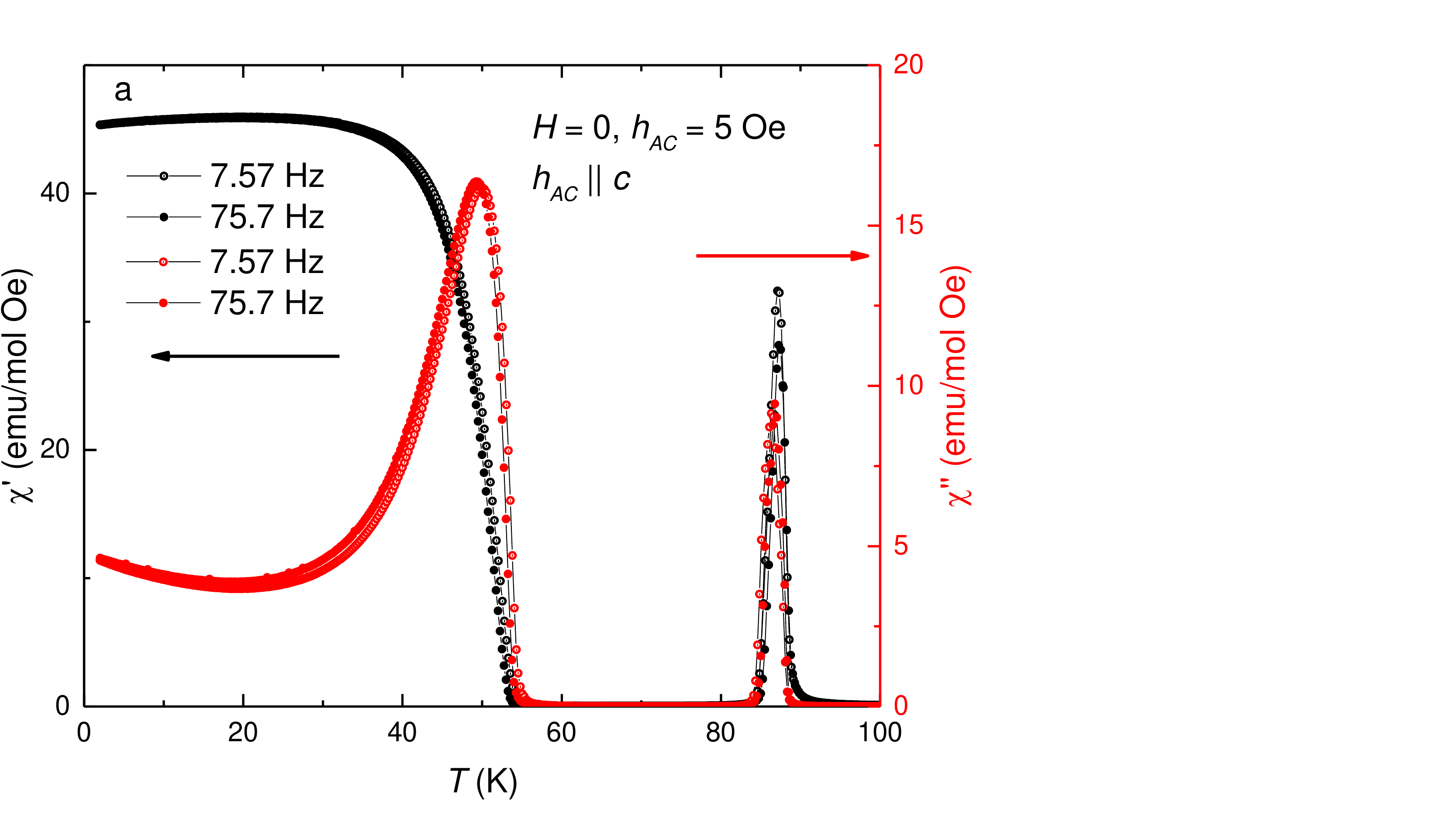}		
	\end{minipage}\hfill
	\centering
	\begin{minipage}{0.45\textwidth}
		\centering
		\includegraphics[width=1.5\columnwidth]{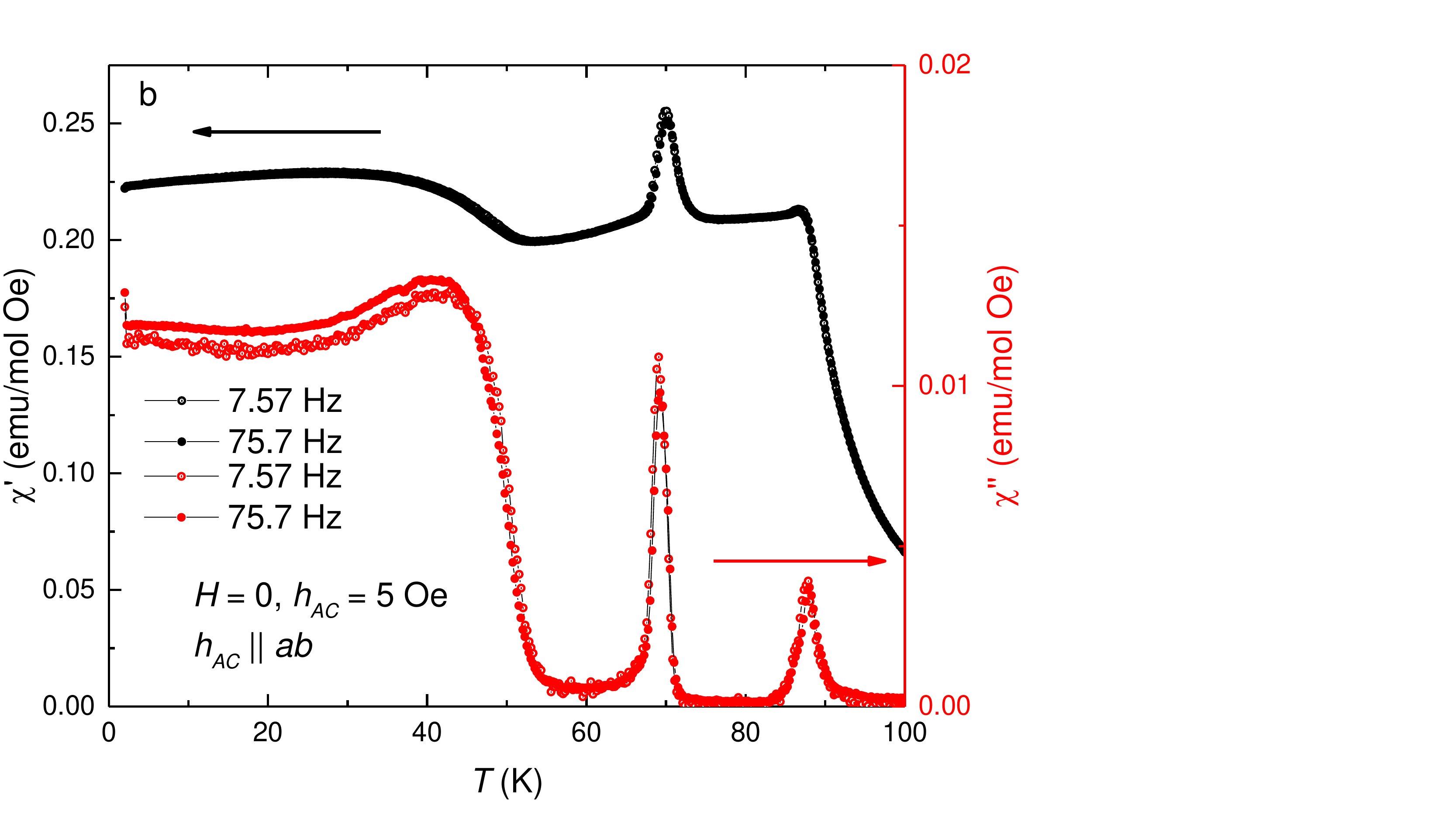}		
	\end{minipage}\hfill
	
	\caption{$\chi^\prime$ and $\chi^{\prime\prime}$ of AC susceptibility as a function of temperature for a LaCrGe$_3$ single crystal with DC field is zero, AC field of 5~Oe, frequency 7.57 Hz and 75.7 Hz. Field is applied parallel (a) or perpendicular (b) to the crystallographic \textit{c}-axis.}\label{22223ACMT}
\end{figure}

Figure \ref{22223ACMT} shows $\chi^\prime$ and $\chi^{\prime\prime}$ of AC susceptibility as a function of temperature for a LaCrGe$_3$ single crystal with zero applied DC field and an AC field of 5~Oe, frequency 7.57 Hz and 75.7 Hz. The AC field is applied parallel (figure \ref{22223ACMT} a) or perpendicular (figure \ref{22223ACMT} b) to the crystallographic \textit{c}-axis. The data manifests very large anisotropy with the scales for the $H$ || $ab$ data sets being two and three orders of magnitude smaller for $\chi^\prime$ and $\chi^{\prime\prime}$ data respectively. The ferromagnetic transition around 86 K is clearly shown. Whereas details of the $H$ || $c$ data will be discussed further below, it is worth pointing out now that, for $H$ || $ab$, the slope of the low field linear $M$($H$) seen below $T_C$ in figure \ref{22223MHdTABF} is the same value as the $\sim$ 0.20 emu/mole-Oe value of $\chi^\prime$ seen below $T_C$ in figure \ref{22223ACMT}b.

\section{discussion and summary}

\begin{figure}
	\includegraphics[width=1.5\columnwidth]{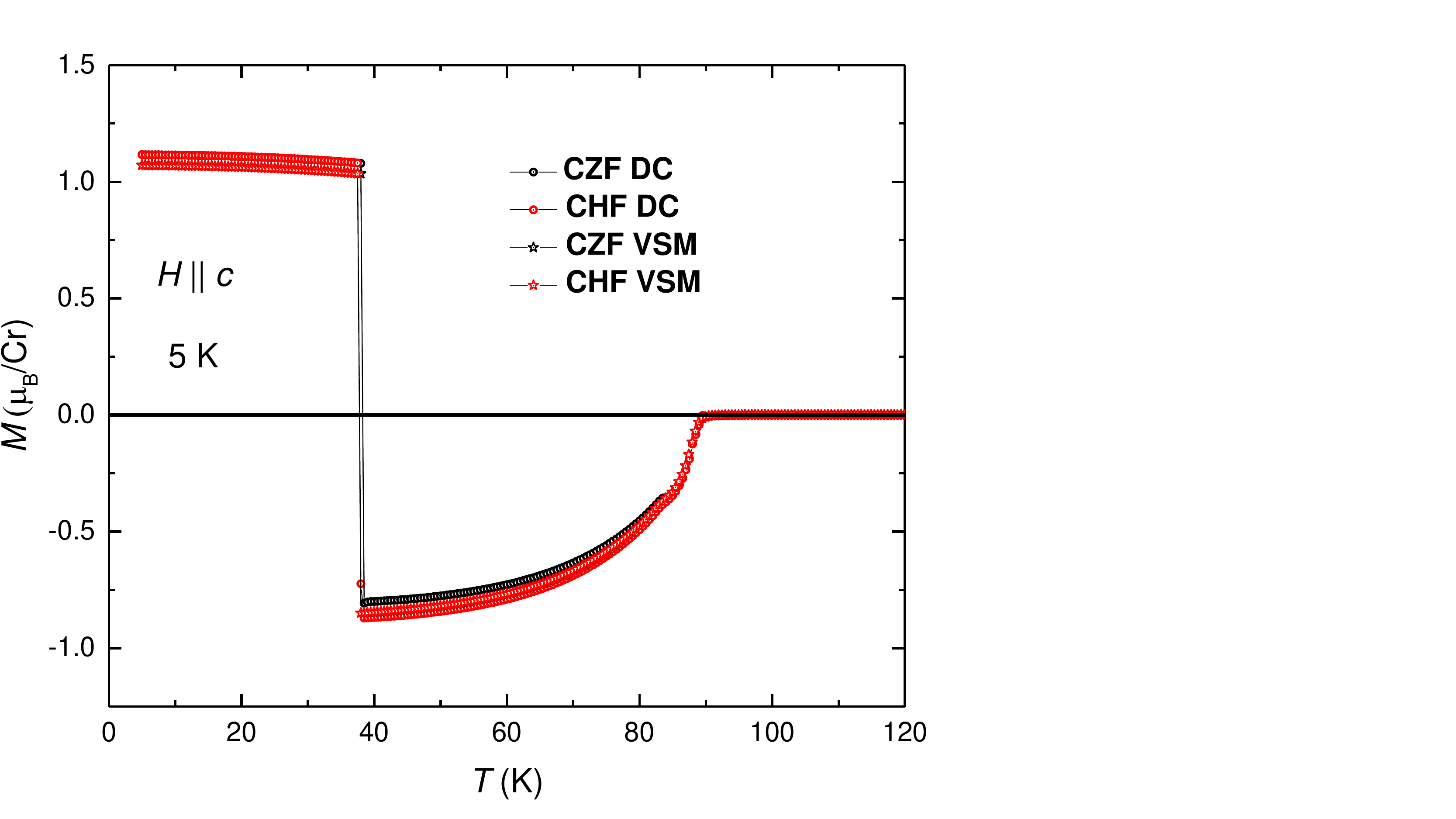}	
	\caption{Cooled in zero field (CZF) and cooled in high field (CHF) magnetization at as a function of temperature (5 K - 120 K) for a LaCrGe$_3$ single crystal after a 20 kOe field applied parallel to the crystallographic \textit{c}-axis was reduced to residual field at 5 K and measured by DC or VSM mode of MPMS3.}\label{22223TrapF}
\end{figure}

The $H$ || $c$ coercivity of single crystalline LaCrGe$_3$, especially for $T$ < 40 K is remarkable. The $M$($H$) loops jump from being fully saturated in the negative-$c$-direction to being fully saturated in the positive-$c$-direction (and vice-versa) in a discontinuous manner. In magnetic nanoparticles with effective magnetic anisotropy constant K (that comes from a variety of sources - shape, magneto-crystalline anisotropy, surface spin canting, etc) and exchange energy per a pair of ions in nanoparticle lattice,  $JS^2$, the width of the domain wall is proportional to $S\sqrt{(J/K)}$, so it becomes thinner in more anisotropic material, but the total surface energy of such a wall is proportional to $S\sqrt{(JK)}$, so both larger exchange constant and larger anisotropy make domain walls energetically less desirable. It seems that in LaCrGe$_3$ we have found a macroscopic bulk system where domain walls are absent and magnetic irreversibility (coercivity) is determined by physics similar to Stoner-Wohlfarth magnetization reversal when the external magnetic field is applied along the easy axis. This is further supported by the fact that the magnetic response in the perpendicular orientation is much smaller. 

The $H$ || $c$ coercivity shown in figure \ref{PD} can be further tested by a variant of what is referred to as a "trapped flux" measurement for superconductors. \cite{Minkov2022a} Figure \ref{22223TrapF} plots the temperature dependence of the magnetization of LaCrGe$_3$ that has either been cooled from 120 K to 5 K in a 20 kOe field (CHF) and then had the field set to zero and warmed or has been cooled from 120 K in zero applied field (CZF) and then at 5 K a field of 20 kOe was applied and subsequently removed. The rate of changing field is 500 Oe/s without overshoot and the interval is 60 s before changing field. In both cases data were taken in the remnant field which we have established to be $\sim$ -19 Oe. Out of an abundance of skepticism we measured the magnetization of the same sample in both a DC mode as well as in a VSM mode All four data sets are shown in figure \ref{22223TrapF} and all four are essentially identical. As the sample is warmed, in a remnant, near zero field, from 5 K its magnetization has a very slight temperature dependence, decreasing from $\sim$1.10 $\mu_B$/Cr to $\sim$1.05 $\mu_B$/Cr as the sample warms through 35 K. Just below 40 K the sample's magnetization jumps, discontinuously to $\sim$ -0.8 $\mu_B$/Cr.  Subsequent warming has the remnant field magnetization decrease monotonically to zero as $T$ increases through $T_C$.  The jump in remnant field magnetization can be understood readily in terms of the coercivity data shown in figure \ref{PD}. As the coercive field decreases from its relatively large value at 5 K toward zero with increasing temperature, it passes through the remnant field value and, at that point, there is the discontinuous change in the samples magnetization. Figure \ref{22223TrapFdF}, in the appendix, shows that we can manipulate the sign and size of the $T$ $\sim$ 40 K jump by adjusting the size and sign of small (or remnant) field experienced by the sample.

The $M$($H$) isotherms demonstrate that there is some change in the details of the magnetic order of LaCrGe$_3$ that results in rather dramatic changes in its coercivity. The AC susceptibility data also indicate that below $T_C$ there is at least one other significant temperature.  For $H$ || $c$, there is a sharp feature at $T_C$ and then near 55 K a second feature. The lower temperature, 55 K, features are most likely associated with the onset of the lower temperature, strongly hysteretic $M$($H$) behavior. Although this occurs at a significantly lower temperature ($\sim$38 K) for the data shown in figures \ref{22223MHdTC}, \ref{PD} and \ref{22223TrapF}, figure \ref{PDd}, in the appendix, shows that for a second sample, with a larger 5 K coercive field, the lower temperature hysteretic region extends to 50 - 55 K.  Given that the jump in magnetization seen in figure \ref{22223TrapF} occurs at the temperature when the remnant field is equal to the coercivity, the larger the 5 K coercivity, the higher the slope of the low temperature part of the coercivity curve (figure \ref{PDd}) and the more accurately this jump determines the temperature of the change in LaCrGe$_3$'s magnetic nature. Both the coercive field data (figures \ref{PD} and \ref{PDd}) as well as the AC susceptibility data (figure \ref{22223ACMT}) support the hypothesis that there is some change in the nature of the ferromagnetically ordered state near 50 - 55 K. The challenge for future work will be to determine the precise nature of this change.

In summary, LaCrGe$_3$ continues to be a fascinating and phenomenal compound. In addition to its avoided quantum criticality, \cite{Taufour2016,Taufour2018} we have now demonstrated that LaCrGe$_3$ has anomalous $M$($H$) curves exhibiting full saturation with sharp transitions and substantial coercivity. The origin of this behavior needs further study. It seems likely that very low bulk pinning, strong uniaxial anisotropy and, possibly, strong surface pinning, are needed to explain the $M$($H$) and $M$($T$) curves. The surface pinning is invoked to explain the observation that the sample needs to be driven rather deeply into the saturated for the full coercivity to manifest (i.e. the need to drive out all domain walls from the bulk sample).  What determines the size of the base temperature coercivity and how sensitive this coercivity is to geometry (i.e length, width, demagnetization factor, etc) or surface roughness or other  details all need to be studied further.  An even bigger question is whether other ferromagnetic materials can be found that demonstrate zero field pinning of fully saturated magnetization over wide temperature ranges and how large of a saturated moment can be stabilize in this manner.  As such LaCrGe$_3$ simply highlights the fact that basic and applied research on intermetallic ferromagnet still has many surprises yet to uncover.

\begin{acknowledgements}
M. Xu thank J. Schmidt and B. Kuthanazhi for valuable discussions. Work at Ames National Laboratory was supported by the U.S. Department of Energy, Office of Science, Basic Energy Sciences, Materials Sciences and Engineering Division. Ames National Laboratory is operated for the U.S. Department of Energy by Iowa State University under contract No. DE-AC02-07CH11358.
\end{acknowledgements}
\clearpage
\renewcommand\refname{[References]}
\bibliographystyle{apsrev}
\bibliography{LaCrGe0303}

\clearpage
\section{Appendix}
\begin{figure}[H]
	\includegraphics[width=1.5\columnwidth]{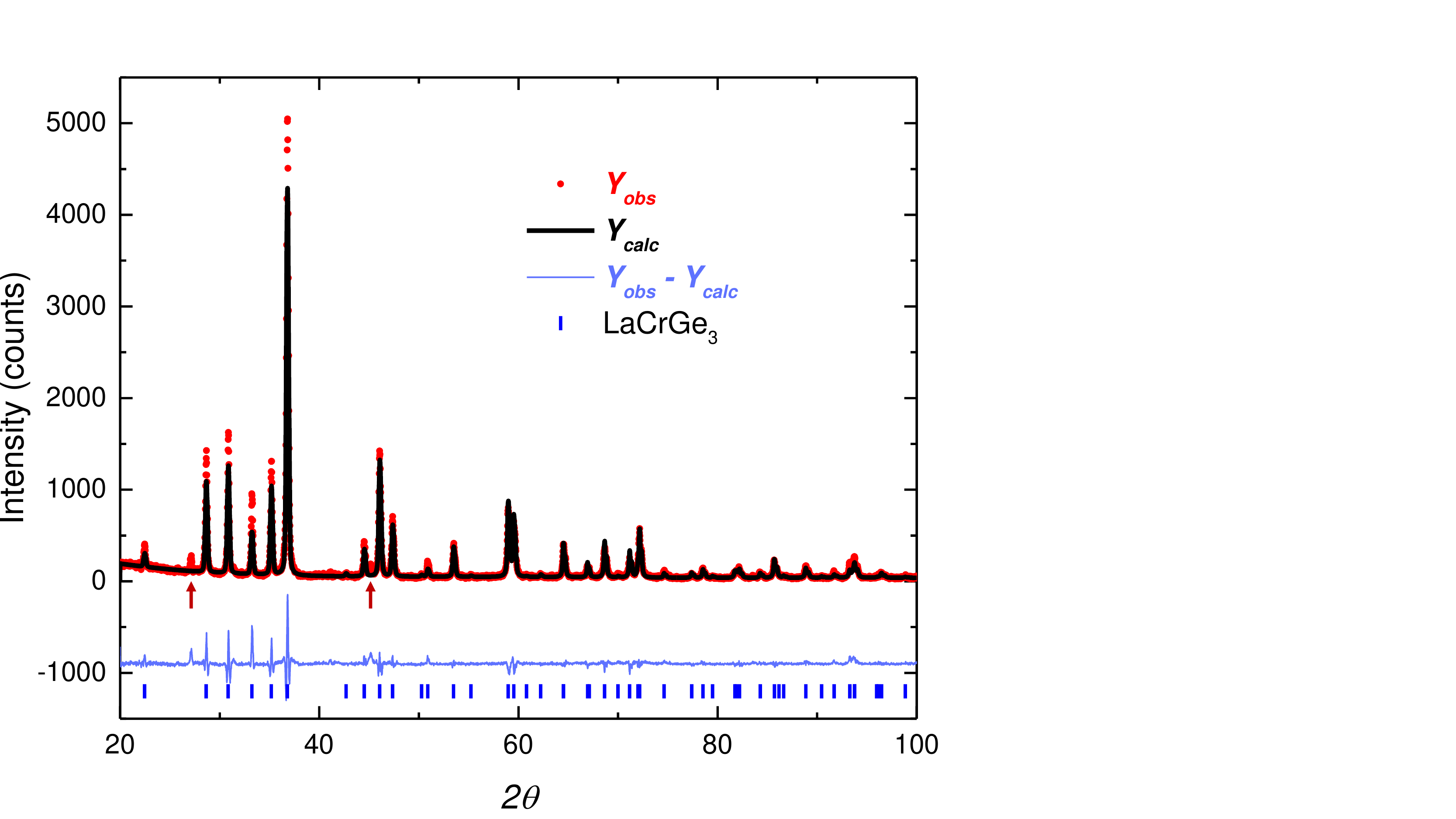}	
	\caption{ Powder X-ray diffraction data of LaCrGe$_3$. Red arrows show the peaks of a small amount of Ge impurity primarily associated with droplets of solidified Ge left over from the decanting process. }\label{22223XRD}
\end{figure}
Figure \ref{22223XRD} presents X-ray diffraction data on a fine powder of formerly single-crystalline LaCrGe$_3$.~Powder X-ray diffraction measurement is carried out to confirm the LaCrGe$_3$ phase as well as identify a very small amount of Ge impurity phase associated with small solidified droplets of residual high temperature liquid that were not completely removed by the decanting process. Measurement is collected by using a Rigaku Miniflex-II diffractometer, with CuK$\alpha$ radiation. Rietveld refinement is obtained using GSAS-II software\cite{Toby2013}.

\begin{figure}[H]
	\centering
	\begin{minipage}{0.44\textwidth}
		\centering
		\includegraphics[width=1.5\columnwidth]{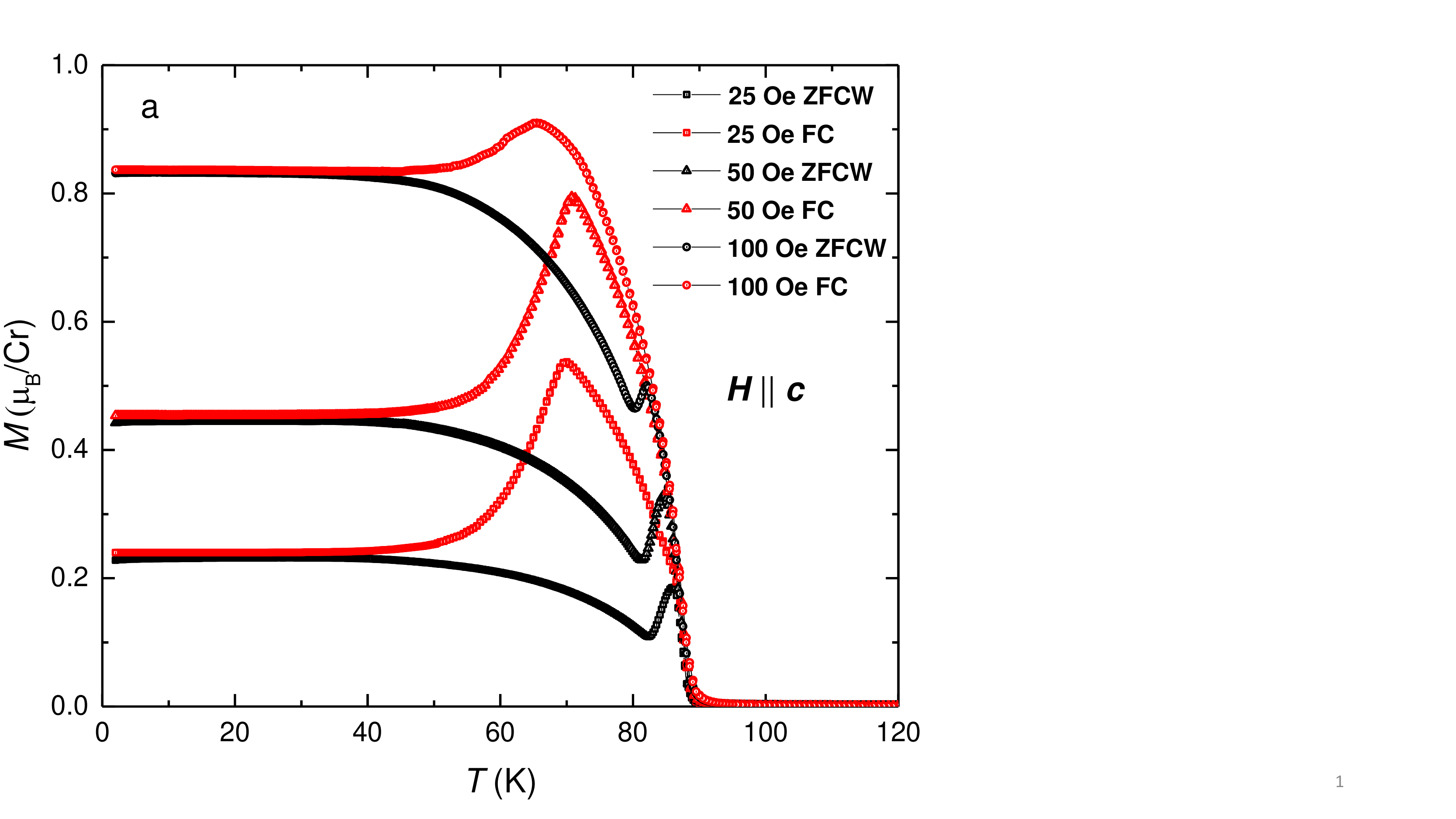}		
	\end{minipage}\hfill
	\centering
	\begin{minipage}{0.44\textwidth}
		\centering
		\includegraphics[width=1.5\columnwidth]{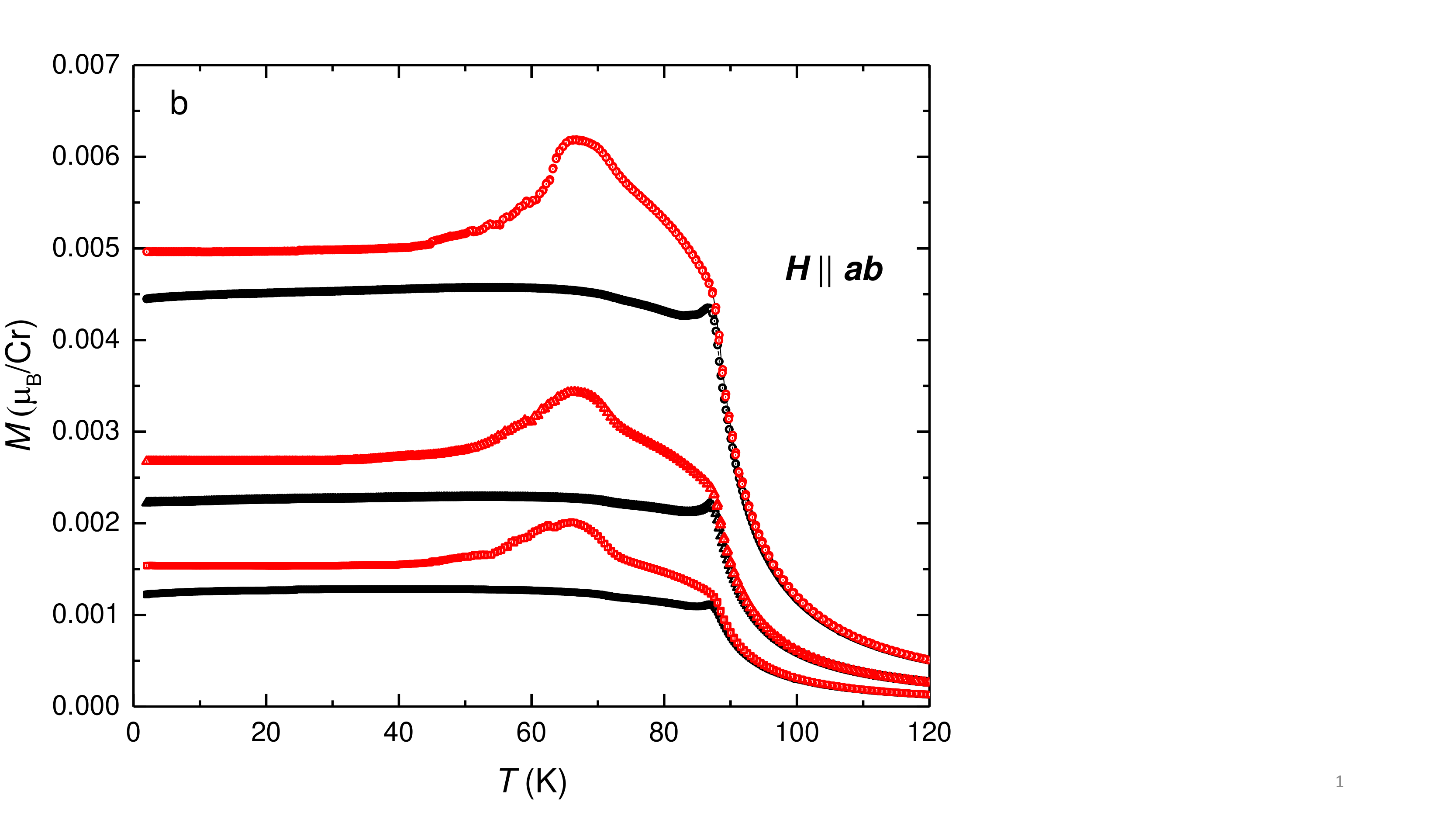}		
	\end{minipage}\hfill
	
	\caption{ZFCW and FC low temperature magnetization as a function of temperature for the LaCrGe$_3$ single crystal with a field of 25, 50 and 100~Oe applied parallel (a) or perpendicular (b) to the crystallographic \textit{c}-axis. }\label{22223MTA}
\end{figure}

Figure \ref{22223MTA} shows the low temperature (1.8~K - 120~K), ZFCW and FC magnetization of LaCrGe$_3$ single crystal with magnetic fields, 25 Oe, 50 Oe and 100 Oe parallel (figure \ref{22223MTA} a) or perpendicular (figure \ref{22223MTA} b) to the crystallographic \textit{c}-axis. The ferromagnetic transition, $\sim$ 86 K, and the peak around 70 K are shown as the same 50 Oe FC data with \cite{Lin2013}. Comparing between figures a and b, almost two orders of magnitude moment larger in $H || c$  than $H || ab$.

\begin{figure}[H]
	\centering
	\includegraphics[width=1.5\columnwidth]{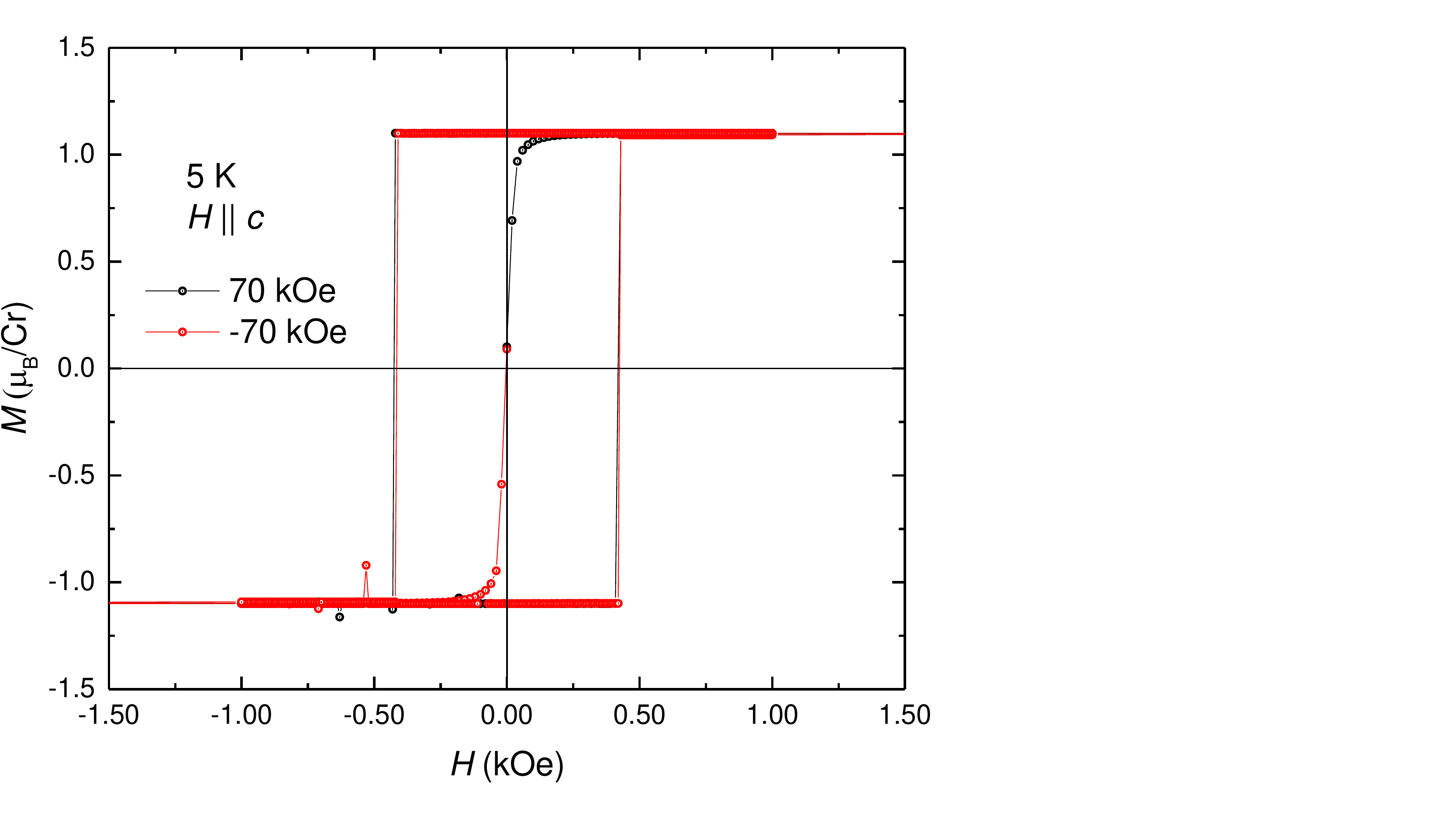}		
	\caption{Magnetization of a single crystal of LaCrGe$_3$ as a function of magnetic field applied parallel to the crystallographic \textit{c}-axis. ZFC to 5 K after demagnetization is done at 120 K before cooling to minimize the remnant magnetic field.  }\label{MH_5K}
\end{figure}

Magnetization as a function of magnetic field applied parallel to the crystallographic \textit{c}-axis at 5 K is shown in figure \ref{MH_5K}.  The black and red symbols presents different directions of initial applied field for the $M$($H$) loops. The direction of the initial field makes no difference on $M$($H$) behavior.

\begin{figure}[H]
	\includegraphics[width=1.5\columnwidth]{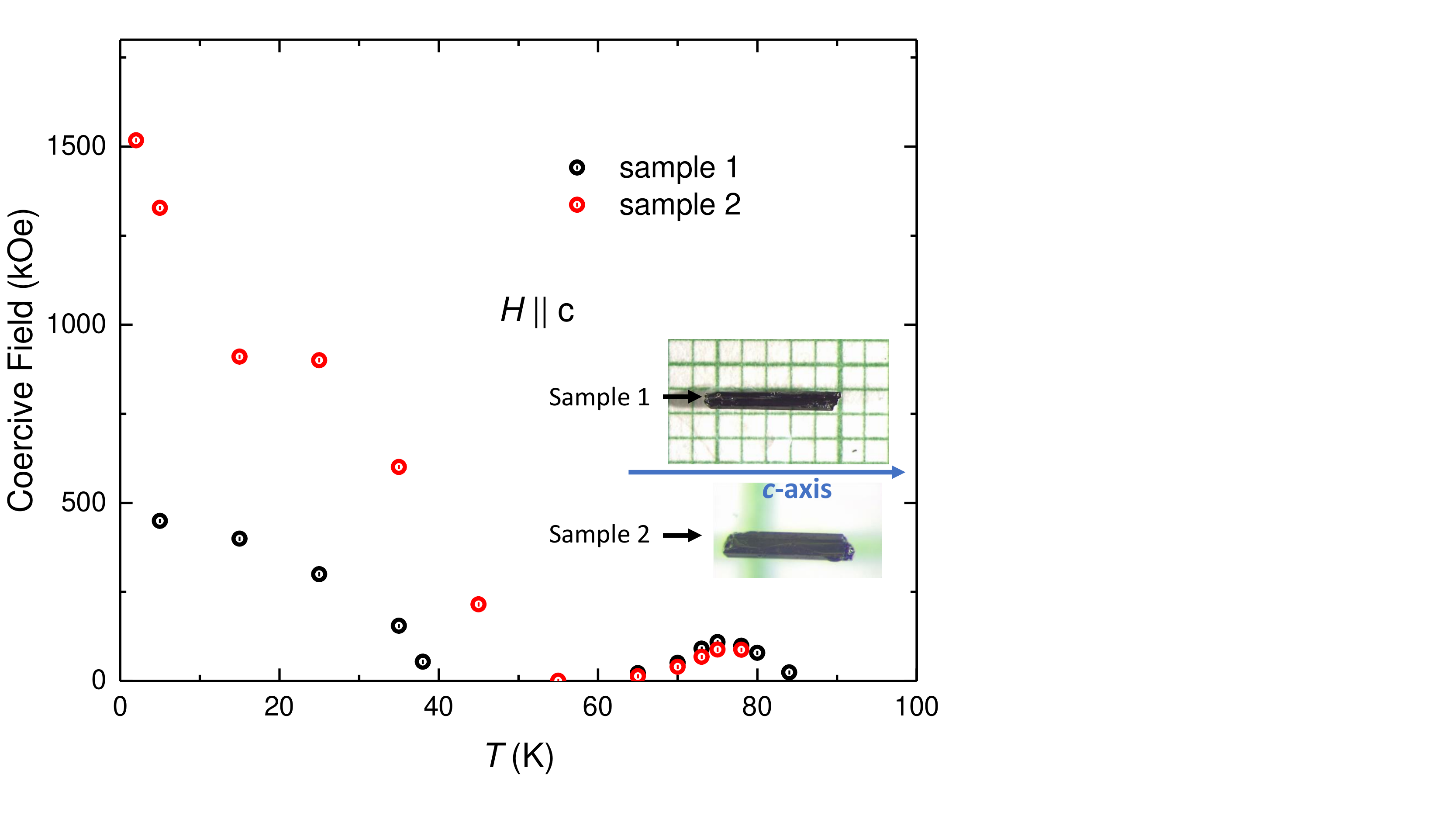}	
	\caption{Coercive field of different samples as a function of temperature with magnetic field applied parallel to the crystallographic \textit{c}-axis. Inset shows the picture of two samples with blue arrow denotes \textit{c}-axis.}\label{PDd}
\end{figure}

Figure \ref{PDd} shows the plot of coercive fields of two different samples as a function of temperature with magnetic field applied parallel to the crystallographic \textit{c}-axis. Inset shows the picture of two samples with blue arrow denotes \textit{c}-axis.
\begin{figure}[H]
	\includegraphics[width=1.5\columnwidth]{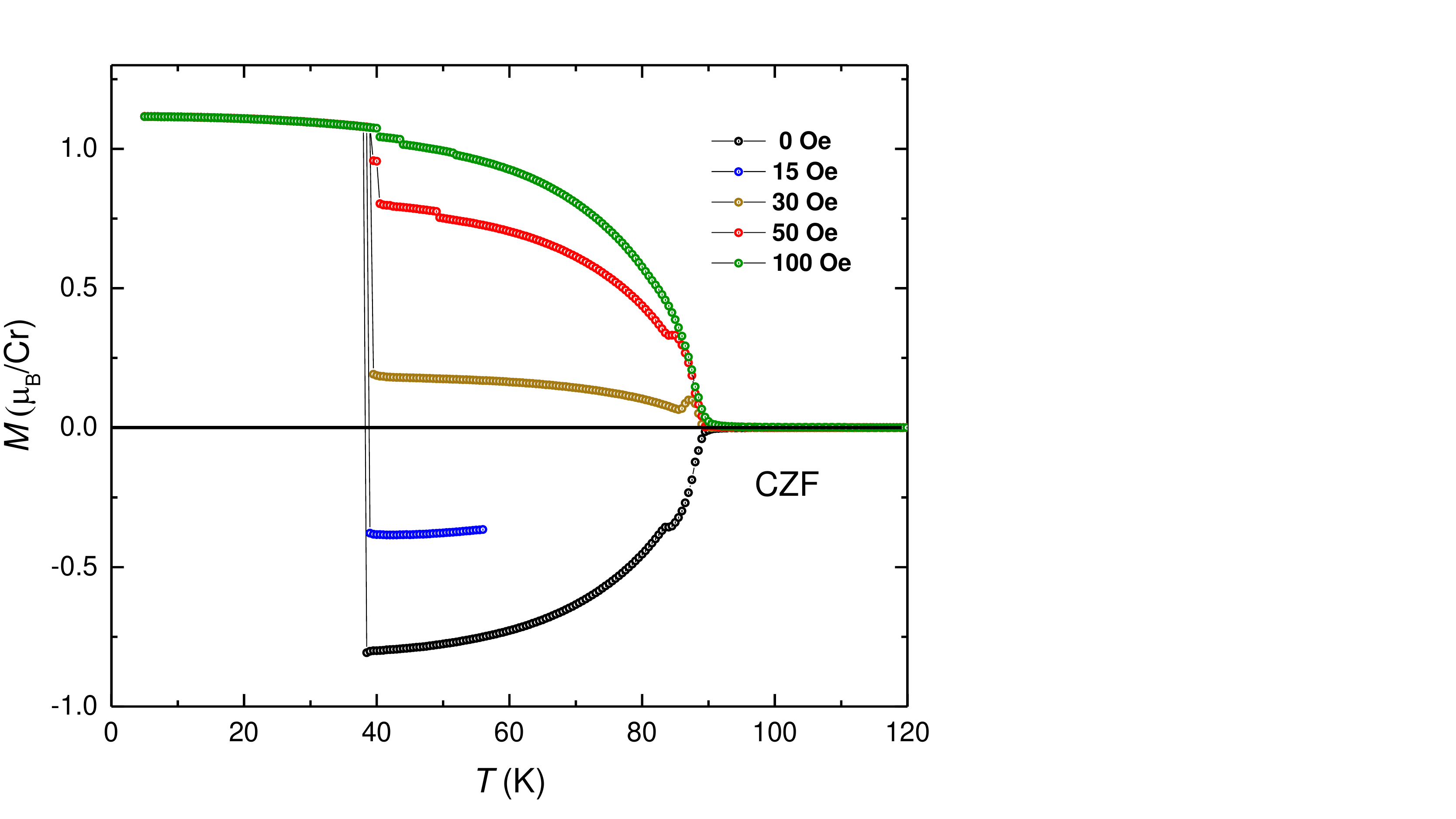}	
	\caption{Magnetization as a function of temperature for single crystalline LaCrGe$_3$ that has been cooled to 5 K in zero field, had a 20 kOe field applied along the $c$-axis, and then had the applied field set to 0, 15, 30, 50, or 100 Oe measured in the MPSM3 unit.  }\label{22223TrapFdF}
\end{figure}
Figure \ref{22223TrapFdF} shows the cooled in zero field (CZF), low temperature magnetization as a function of temperature (5 K - 120 K). The single crystalline LaCrGe$_3$ was cooled to 5 K in zero field, had a 20 kOe field applied along the $c$-axis, and then had the applied field set to 0, 15, 30, 50, or 100 Oe measured in the MPSM3 unit. The value of the moment after the jump is related to the final field applied, but the temperature of jump is almost not affected by applied field. 

\end{document}